# Rayleigh scattering under light-atom coherent interaction

Akifumi Takamizawa[1*] and Koichi Shimoda[2†]

[1]*National Metrology Institute of Japan (NMIJ), National Institute of Advanced Industrial Science and Technology, Tsukuba, Ibaraki 305-8563, Japan*
[2]*The Japan Academy, Ueno Park, Tokyo 110-0007, Japan*

Semi-classical calculation of an oscillating dipole induced in a two-level atom indicates that spherical radiation from the dipole under coherent interaction, i.e., Rayleigh scattering, has a power level comparable to that of spontaneous emission resulting from an incoherent process. Whereas spontaneous emission is nearly isotropic and has random polarization generally, Rayleigh scattering is strongly anisotropic and polarized in association with incident light. In the case where Rabi frequency is much larger than frequency detuning and spontaneous emission rate, while the power of spontaneous emission varies with time at the Rabi frequency, the power of Rayleigh scattering oscillates at twice the Rabi frequency. Moreover, the radiation pressure force acting on an atom due to Rayleigh scattering exceeds that caused by spontaneous emission when frequency detuning is large.

## 1. Introduction

The directionality of stimulated emission and spontaneous emission plays important roles in optical science and atomic physics. For example, a laser cavity can be composed of two mirrors facing each other, since the stimulated emission light contributing to coherent amplification appears in a direction identical to that of the incident light. In contrast, the photon scattering force acting on an atom due to spontaneous emission plays an essential role in techniques for laser cooling of atoms such as Doppler cooling [1, 2] and polarization gradient cooling [3].

Both types of emission are generated by spherical radiation from an oscillating dipole induced in an atom. Here, stimulated emission results from a dipole coherently oscillating with incident light. On the other hand, spontaneous emission is generated by a dipole oscillating at the atomic resonant frequency without phase correlation to the incident light when an atom in an excited state spontaneously decays to a ground state. Consideration of interference between coherent spherical radiation and incident light can explain the directionality of the stimulated emission [4, 5].

However, it should be considered that a term representing spherical radiation accompanies the interference term. Incident light is thereby dissipated by the coherently oscillating dipole. Photon recoil is caused by this light scattering (i.e., Rayleigh scattering), and the atom is subjected to radiation pressure force. Namely, light scattering and radiation pressure force are not caused only by incoherent spontaneous emission.

Since 1969, it has been known that the spectrum of light dissipated by a two-level atom has a triplet structure called Mollow triplet [6-10]. The spectral structure results from spontaneous emission, Rayleigh scattering and Raman scattering [7, 8]. Recently, however, in contrast to spontaneous emission and Raman scattering, whose power is the same as that of spontaneous emission [7], Rayleigh scattering has been downgraded in the field of atomic physics and in its application except for the superradiant Rayleigh scattering from a Bose-Einstein condensate [11, 12]. For example, radiation pressure force has been usually explained by variation in the momentum of photons which are absorbed and scattered due to spontaneous emission [13]: Rayleigh scattering has not been taken into account.

In this paper, Rayleigh scattering caused by coherent interaction between incident light and a two-level atom is derived semi-classically. Although spontaneous emission is not included in the calculation, we can regard light as electromagnetic waves in contrast to full quantum calculation in which light is treated as photons. The semi-classical approach enables us to intuitively find that Rayleigh scattering is inevitable under coherent interaction. Moreover, we can simply and straightforwardly obtain not only the power of Rayleigh scattering but also its time dependence. Consequently, it is shown that Rayleigh scattering is not negligible compared to spontaneous emission, whose photon number per unit time is derived by multiplying the population of the excited state by the spontaneous emission rate, in terms of power and radiation pressure force acting on an atom. Note that we here describe Rayleigh scattering from an ensemble of uncorrelated atoms.

The power of Rayleigh scattering is calculated in §2. Then, the differences in processes, time dependence and coherence between Rayleigh scattering and spontaneous emission are shown in §3. In addition, the effect of Rayleigh scattering on atom detection is discussed, and radiation pressure force acting on an atom due to Rayleigh scattering

[*] email: akifumi.takamizawa@aist.go.jp
[†] professor emeritus at The University of Tokyo

is compared to that due to spontaneous emission. Finally, we give a conclusion in §4.

## 2. Calculation of the power of Rayleigh scattering

For theoretical calculation, it is assumed that a two-level atom with a wavefunction $\psi = a_1(t)\psi_1 + a_2(t)\psi_2$ ($t$: time) is in a vacuum, where $\psi_1$ and $\psi_2$ are the eigenfunctions of the ground state and the excited state, respectively. The origin of an inertial Cartesian coordinate ($x$, $y$, $z$) is taken at the position of the atom. Incident light whose electric field is expressed as $\boldsymbol{E}_\mathrm{L} = E_0 \cos(\omega t - kx)\hat{z}$ ($\hat{z}$: a unit vector in the $z$ direction) is incident to the atom, where $E_0$, $\omega$ and $k$ are the amplitude of the electric field, the frequency and the wave number, respectively.

Here, we consider a simple case only with coherent interaction, neglecting spontaneous emission. Under coherent interaction caused by the perturbation Hamiltonian $H' = -\mu_z E_\mathrm{L}$ ($\mu_z$: $z$ component of the electric dipole moment), the population in the ground state and that in the excited state are given by $|a_1(t)|^2 = 1 - (\Omega_0/\Omega)^2 \sin^2(\Omega t/2)$ and $|a_2(t)|^2 = 1 - |a_1(t)|^2$, respectively, where the initial conditions $a_1(0) = 1$ and $a_2(0) = 0$ are taken. Here, $\Omega = (\Delta^2 + \Omega_0^2)^{1/2}$ is the Rabi frequency, where $\Delta = \omega - \omega_0 + \delta_\mathrm{D}$ is the frequency detuning of the incident light with respect to the atomic resonant frequency $\omega_0$. Here, $\delta_\mathrm{D}$ is the Doppler shift caused by the movement of the atom with respect to the light source. The vacuum Rabi frequency is $\Omega_0 = |\mu_{21} E_0|/\hbar$, where $\mu_{21}$ and $2\pi\hbar$ are the matrix element of $\mu_z$ and Planck's constant $h$, respectively. The quantum-mechanical expectation value of the induced dipole $\int \psi^* ez\psi \, d\boldsymbol{r}$ ($e$: the electron charge) is given by $\boldsymbol{p}(t) = p_0(t)\exp(i\omega t)\hat{z} + c.c.$, where [5]

$$p_0(t) = -\frac{\Omega_0}{2\Omega}|\mu_{21}|\left[\frac{\Delta}{\Omega}(1-\cos\Omega t) + i\sin\Omega t\right]. \quad (1)$$

The light intensity is composed of the incident light and the radiation from the induced dipole as given by $|\boldsymbol{E}|^2 = |\boldsymbol{E}_\mathrm{L}|^2 + |\boldsymbol{E}_\mathrm{P}|^2 + 2\boldsymbol{E}_\mathrm{L} \cdot \boldsymbol{E}_\mathrm{P}$, where $\boldsymbol{E}_\mathrm{P}$ is the electric field of the radiation from the induced dipole. When $\omega \gg \Omega$ and $kR \gg 1$, the intensity is given by

$$|\boldsymbol{E}_\mathrm{P}|^2 = \frac{k^4 \sin^2\theta}{8\pi^2 \varepsilon_0^2 R^2}|p_0(t)|^2 \quad (2)$$

in a spherical coordinate ($R$, $\theta$, $\varphi$), where $R = (x^2 + y^2 + z^2)^{1/2}$ and $\theta = \cos^{-1}(z/R)$.

Here, $|\boldsymbol{E}_\mathrm{P}|^2$ can be regarded as the intensity caused by Rayleigh scattering [14]. On the other hand, the interference term $2\boldsymbol{E}_\mathrm{L} \cdot \boldsymbol{E}_\mathrm{P}$ expresses the change in intensity caused by absorption, stimulated emission and dispersion. The spatial distribution of the interference term will be described separately [15].

While the statistically and temporally averaged intensity of Rayleigh scattering is generally calculated for small particles, the time dependence of the intensity of Rayleigh scattering from an individual atom is given by Eqs. (1) and (2), which are derived from quantum mechanical calculation. By integrating the intensity over the whole solid angle, the power of Rayleigh scattering from the two-level atom is obtained as

$$P_\mathrm{ray} = \frac{\hbar\omega\Gamma}{4} \cdot \frac{\Omega_0^2}{\Omega^2}\left[\frac{\Delta^2}{\Omega^2}(1-\cos\Omega t)^2 + \sin^2\Omega t\right], \quad (3)$$

where $\Gamma$ is the spontaneous emission rate, and $\omega \cong \omega_0$ is taken assuming $|\Delta| \ll \omega$. The time average of $P_\mathrm{ray}$ over the Rabi oscillation is given by

$$\overline{P_\mathrm{ray}} = \frac{\hbar\omega\Gamma}{8} \cdot \frac{\Omega_0^2}{\Omega^2}\left(1 + \frac{3\Delta^2}{\Omega^2}\right). \quad (4)$$

## 3. Discussion

### 3.1 Differences between Rayleigh scattering and spontaneous emission

Rayleigh scattering is similar to spontaneous emission in that both are spherical radiation. However, Rayleigh scattering is different from spontaneous emission, because Rayleigh scattering can be explained by light-atom coherent interaction whereas spontaneous emission cannot [13].

Moreover, Rayleigh scattering makes no variation in the internal energy of an atom in contrast to spontaneous emission. The reason is described as follows: Under coherent interaction, the internal energy of an atom varies by the work which is given by the light-atom interaction. The variation in the internal energy corresponds to absorption or stimulated emission. Here, since the work per unit time is given by $-\omega E_0 \,\mathrm{Im}[p_0(t)]$, the phase difference between incident light and an induced dipole is very important for variation in the internal energy. However, the intensity of Rayleigh scattering is not dependent on the phase of an induced dipole but its amplitude as shown in Eq. (2).

Let us discuss differences in time dependence of radiation power between Rayleigh scattering and spontaneous emission. The power of spontaneous emission is given by $P_\mathrm{spo} = \hbar\omega_0\Gamma|a_2(t)|^2$, whereas the power of Rayleigh scattering is proportional to $|p_0(t)|^2$, as shown in Eq. (2). Figures 1(a) and 1(b) show $|a_2(t)|^2$ and $-\mathrm{Im}[p_0(t)]$, respectively, as a function of $\Omega t$ for $\Omega \gg |\Delta|$ and $\Omega \gg \Gamma$. Here, note that $|p_0(t)|^2 = |\mathrm{Im}[p_0(t)]|^2$ for $\Omega \gg |\Delta|$, and that temporal variations in $|a_2(t)|^2$ and $\boldsymbol{p}(t)$ caused by spontaneous emission are negligible for $\Omega \gg \Gamma$. Since the power of the absorbed light is obtained as $-\omega E_0 \text{Im}[p_0(t)]$, $|a_2(t)|^2$ and $\text{Im}[p_0(t)]$ are connected by the equation $\hbar\omega_0|a_2(t)|^2 = -\omega E_0 \int_0^t \text{Im}[p_0(t)]dt$. From this equation, the relationship between $P_{\text{ray}}$ and $P_{\text{spo}}$ can be explained using $\text{Im}[p_0(t)]$: $P_{\text{ray}} \propto |\text{Im}[p_0(t)]|^2$ and $P_{\text{spo}} \propto -\int_0^t \text{Im}[p_0(t)]dt$.

Figure 1(c) shows $P_{\text{ray}}$ and $P_{\text{spo}}$ with solid and broken lines, respectively, as a function of $\Omega t$ for $\Omega \gg |\Delta|$ and $\Omega \gg \Gamma$. As shown in Fig. 1(c), $P_{\text{ray}}$ oscillates at twice the Rabi frequency, while $P_{\text{spo}}$ varies with time at the Rabi frequency. This should be useful for discrimination between Rayleigh scattering and spontaneous emission by measurement of radiation. However, this is not the case when $\Omega \cong |\Delta|$ at large detuning, since both Rayleigh scattering and spontaneous emission are proportional to $\sin^2(\Omega t/2)$.

In addition, it is found from Fig. 1(c) that $\overline{P_{\text{spo}}} = 4\overline{P_{\text{ray}}}$, where $\overline{P_{\text{spo}}}$ is the time average of $P_{\text{spo}}$ over the Rabi oscillation, so the power of Rayleigh scattering is comparable to that of spontaneous emission.

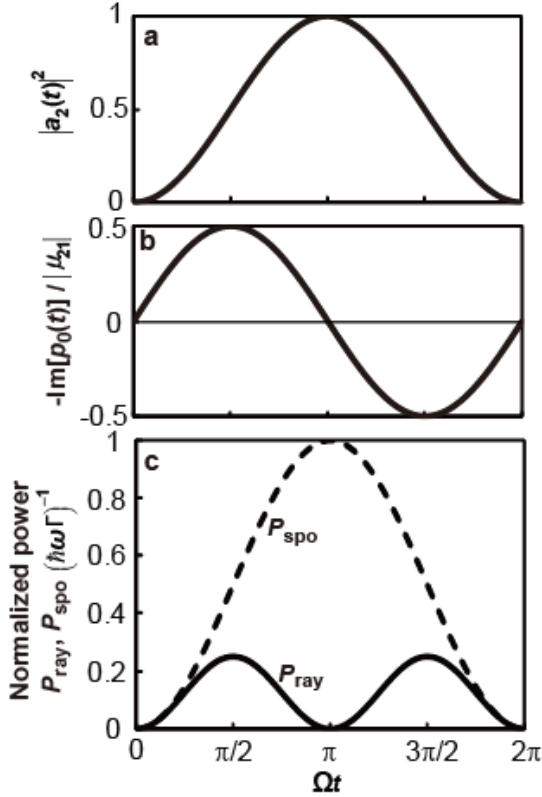

Fig. 1. (a) Population in an excited state, $|a_2(t)|^2$, and (b) $-\text{Im}[p_0(t)]$ as a function of $\Omega t$, where $p_0(t)$ is the amplitude of the induced dipole. (c) The power of Rayleigh scattering $P_{\text{ray}}$ and that of spontaneous emission $P_{\text{spo}}$ are shown as a function of $\Omega t$ with solid and broken lines, respectively. Here, we assume $\Omega \gg |\Delta|$ and $\Omega \gg \Gamma$. Also, $-\text{Im}[p_0(t)]$ is normalized by $|\mu_{21}|$, whereas $P_{\text{ray}}$ and $P_{\text{spo}}$ are normalized by $\hbar\omega\Gamma$. Here, $P_{\text{ray}}$ and $P_{\text{spo}}$ are proportional to $|\text{Im}[p_0(t)]|^2$ and $|a_2(t)|^2$, respectively. As a result, $P_{\text{ray}}$ varies with $\sin 2\Omega t$, whereas $P_{\text{spo}}$ changes with $\sin\Omega t$. In terms of the time-averaged power, we find that $\overline{P_{\text{spo}}} = 4\overline{P_{\text{ray}}}$.

If the condition $\Omega \gg \Gamma$ is not satisfied, the changes in $|a_2(t)|^2$ and $p(t)$ caused by spontaneous emission are not negligible. However, if $\Omega \gtrsim \Gamma$, the power and its time-average for Rayleigh scattering can be roughly estimated using Eqs. (3) and (4), respectively.

In terms of the radiation field, Rayleigh scattering and spontaneous emission can be distinguished by their coherence. Rayleigh scattering is coherent with incident light, and their frequencies are identical. Furthermore, when laser light is incident to an atomic ensemble, Rayleigh scattering from individual atoms causes interference. Moreover, since the induced dipole oscillates in the incident light's direction of polarization, Rayleigh scattering is strongly anisotropic and polarized.

On the other hand, spontaneous emission is incoherent since it has no phase relation to incident light [16]. Moreover, under the condition that $|\Delta| \gg \Omega_0$, the frequency of spontaneous emission is identical to the atomic resonant frequency [7-9]. Therefore, Rayleigh scattering can be distinguished from spontaneous emission by their frequency difference at large detuning and low intensity of incident light. (Even when $|\Delta| \lesssim \Omega_0$, either of the side peaks of a Mollow triplet structure corresponds to the atomic resonant frequency deviated by a light shift $\delta_{\text{ls}} = \Delta[1-(\Omega/|\Delta|)]$ [7]. However, we cannot strictly determine which peak corresponds to the frequency of spontaneous emission for $|\Delta| \lesssim \Omega_0$ since it is necessary to consider a transition between dressed states, which are expressed by superposition between uncoupled states [8].) Additionally, the polarization of spontaneously emitted light is determined by variation in the magnetic quantum number in the transition, $\Delta m_{\text{spo}}$: Spontaneously emitted light has $\sigma^-$-, $\pi$- and $\sigma^+$-polarizations if $\Delta m_{\text{spo}} = +1$, 0 and $-1$, respectively. If $\Delta m_{\text{spo}} = -\Delta m_{\text{abs}}$, where $\Delta m_{\text{abs}}$ is variation in the magnetic quantum number when an atom absorbs incident light, the polarization and the directionality of spontaneous emission are identical to those of Rayleigh scattering. In general, spontaneous emission is nearly isotropic with random polarization in contrast to Rayleigh scattering since most kinds of atoms have a few or more Zeeman sublevels in hyperfine levels of the ground state [13]. However, in the case of an atom with a nondegenerated ground state such as an alkaline-earth metal atom, spontaneous emission cannot be discriminated from Rayleigh scattering by polarization and directionality.

### 3.2 Atom detection

Let us consider a case where fluorescence induced by resonant light is observed. Rayleigh scattered light and spontaneously emitted light are simultaneously received by a photodetector. However, when the number of atoms increases, the relative intensity of Rayleigh scattering with respect to the total intensity of the dissipated light decreases due to interference between the individual dipoles: Rayleigh scattering from each atom interferes constructively in the direction parallel to that of incident light but destructively in other directions.

In the case of a single atom [17-20], the Rayleigh scattering intensity is purely obtained as $|E_p|^2$. As seen from Eq. (2), the Rayleigh scattering intensity is zero in the direction $\theta = 0$, and is maximized in the direction $\theta = \pi/2$. The maximum intensity is calculated from Eq. (2) as 3/2 times the intensity averaged over the whole solid angle. Accordingly, for $\Omega \gg |\Delta|$ and $\Omega \gg \Gamma$, in which $\overline{P_{spo}} = 4\overline{P_{ray}}$, the Rayleigh scattering intensity in the direction $\theta = \pi/2$ is 3/8 times the intensity of spontaneous emission assuming isotropic spontaneous emission. Hence, the light scattered from a single atom can be efficiently observed by optimizing the location of the photodetector with respect to the polarization of the incident light.

### 3.3 Radiation pressure force

Let us discuss the radiation pressure force acting on an atom due to Rayleigh scattering. Since the statistically averaged momentum of photons in Rayleigh scattering is zero because of the symmetry for $\theta$, radiation pressure force caused by Rayleigh scattering per unit time is given by $F_{ray} = P_{ray}/c$, where $c (:= \omega/k)$ is the speed of light in the vacuum. The strength of the radiation pressure force therefore oscillates as shown in Eq. (3). The time average of $F_{ray}$ over the Rabi oscillation can be obtained as $\overline{F_{ray}} = \overline{P_{ray}}/c$ by using $\overline{P_{ray}}$ in Eq. (4). Here, note that the condition $\Omega \gtrsim \Gamma$ must be satisfied to express $F_{ray}$ and $\overline{F_{ray}}$ from Eqs. (3) and (4), respectively. On the other hand, radiation pressure force caused by spontaneous emission per unit time is given by $F_{spo} = \hbar k \Gamma s [2(1+s)]^{-1}$ in a steady state [13]. Here, $s = (I/I_s)[1 + (2\Delta/\Gamma)^2]^{-1}$ is the saturation parameter, where $I$ and $I_s$ are the intensity of the incident light and the saturation intensity, respectively. Using $I$ and $I_s$, the vacuum Rabi frequency is given by $\Omega_0 = [I/(2I_s)]^{1/2} \Gamma$.

Figure 2(a) shows $\overline{F_{ray}}$ and $F_{spo}$ at $|\Delta|/\Gamma = 1$ as a function of $I$ with solid and broken lines, respectively. Figure 2(b) represents $\overline{F_{ray}}$ and $F_{spo}$ at $I/I_s = 10$ as a function of $|\Delta|$ with solid and broken lines, respectively. Here, $\Omega \geq \Gamma$ is satisfied at $|\Delta|/\Gamma \geq 1$ or $I/I_s \geq 10$. Moreover, Fig. 3(a) shows $\overline{F_{ray}}$ for various $|\Delta|$ as a function of $I$, and Fig. 3(b) provides $\overline{F_{ray}}$ for various $I$ as a function of $|\Delta|$. As shown in Figs. 2(a) and 2(b), while $F_{spo}$ simply rises with increasing values of $I$ and decreasing values of $|\Delta|$, $\overline{F_{ray}}$ has a maximum for both $I$ and $|\Delta|$. The maximum of $\overline{F_{ray}}$ is analytically derived from Eq. (4) as $\hbar k \Gamma / 6$ at $I/I_s = 4(\Delta/\Gamma)^2$. Note that the maximum value is independent of $I$ and $\Delta$ as shown in Figs. 3(a) and 3(b).

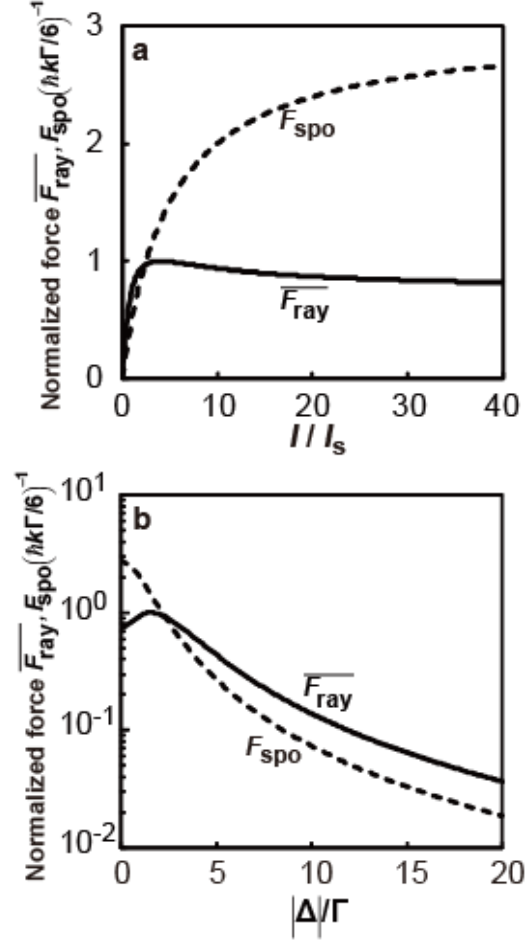

Fig. 2. Radiation pressure force caused by Rayleigh scattering, $\overline{F_{ray}}$, and that caused by spontaneous emission, $F_{spo}$, are shown with solid and broken lines, respectively, as functions of (a) the intensity of incident light, $I$, at $|\Delta|/\Gamma = 1$, and (b) the frequency detuning $|\Delta|$ at $I/I_s = 10$. Here, $\overline{F_{ray}}$ and $F_{spo}$ are normalized by $\hbar k \Gamma / 6$. Also, $I$ and $|\Delta|$ are normalized by $I_s$ and $\Gamma$, respectively. The vertical axis in (b) is logarithmic. As shown in (a), $\overline{F_{ray}}$ is larger than $F_{spo}$ when $I$ is small, but is smaller than $F_{spo}$ when $I$ is large due to rapid saturation. As represented in (b), while $F_{spo}$ monotonically decreases as $|\Delta|$ increases, $\overline{F_{ray}}$ has a peak at $|\Delta|/\Gamma = [I/(4I_s)]^{1/2}$. In addition, we find that $\overline{F_{ray}}$ is smaller than $F_{spo}$ around $|\Delta| = 0$ but exceeds $F_{spo}$ when $|\Delta|$ is large.

As indicated in Fig. 2(a), while $\overline{F_{ray}}$ increases more rapidly than $F_{spo}$ with $I$ for small values of $I$, $\overline{F_{ray}}$ becomes smaller than $F_{spo}$ for large values. It can be analytically derived from Eq. (4) that these observations remain true independently of $\Delta$: It is found that $\overline{F_{ray}}$ is larger than $F_{spo}$ for $I/I_s \ll (\Delta/\Gamma)^2$ and $I/I_s \ll 1$, whereas $\overline{F_{ray}} \cong F_{spo}/4 \cong \hbar k \Gamma / 8$ for $I/I_s \gg (\Delta/\Gamma)^2$ and $I/I_s \gg 1$. Figure 2(b) also shows that $\overline{F_{ray}}$ is smaller than $F_{spo}$ when $|\Delta|$ is small, while $\overline{F_{ray}}$ is larger than $F_{spo}$ when $|\Delta|$ is large. It is analytically found from Eq. (4) that these observations hold independently of $I$: For $|\Delta| \ll \Gamma$, it is calculated as $\overline{F_{\text{ray}}} \cong \hbar k\Gamma/8$, whereas $F_{\text{spo}} > \hbar k\Gamma/3$ under the condition $\Omega \geq \Gamma$. In addition, it is found that $\overline{F_{\text{ray}}} \cong 2F_{\text{spo}} \cong (\hbar k\Gamma/4)(I/I_s)(\Delta/\Gamma)^{-2}$ for $(\Delta/\Gamma)^2 \gg I/I_s$ and $(\Delta/\Gamma)^2 \gg 1$.

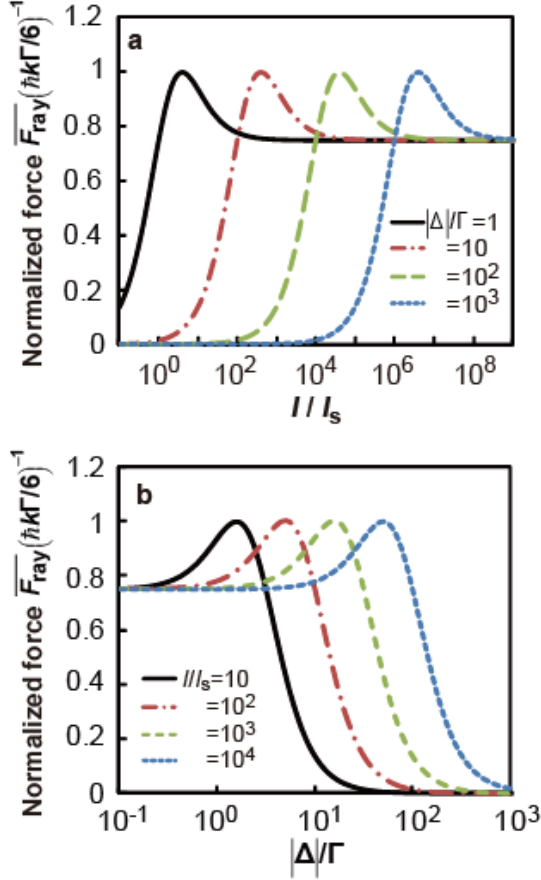

Fig. 3. Radiation pressure force caused by Rayleigh scattering, $\overline{F_{\text{ray}}}$, as functions of (a) the intensity of the incident light, $I$, at $|\Delta|/\Gamma = 1, 10, 10^2$ and $10^3$, and (b) the frequency detuning $|\Delta|$ at $I/I_s = 10, 10^2, 10^3$ and $10^4$. Here, $\overline{F_{\text{ray}}}$ is normalized by $\hbar k\Gamma/6$. Also, $I$ and $|\Delta|$ are normalized by $I_s$ and $\Gamma$, respectively. The horizontal axes are logarithmic. $\overline{F_{\text{ray}}}$ is maximized as $\hbar k\Gamma/6$ at $I/I_s = 4(\Delta/\Gamma)^2$. Together with increasing values of $I$ or decreasing values of $|\Delta|$, $\overline{F_{\text{ray}}}$ approaches a constant value of $\hbar k\Gamma/8$.

The condition $\Omega \gg \Gamma$ is not always satisfied for $I \ll I_s$, but is inevitably fulfilled for $|\Delta| \gg \Gamma$. It is therefore certain that $\overline{F_{\text{ray}}}$ is larger than $F_{\text{spo}}$ for $|\Delta| \gg \Gamma$. Hence, the radiation pressure force caused by Rayleigh scattering must be especially significant for far-detuned-light atom manipulation such as that seen in a far-off resonance trap [21] and atom reflection with blue-detuned evanescent light [22-25].

## 4. Conclusion

We find that spherical radiation from the induced dipole exists even under coherent interaction, and can be regarded as Rayleigh scattering. Although Rayleigh scattering is different from spontaneous emission in terms of coherence and time dependence, it is comparable to spontaneous emission for $\Omega \gtrsim \Gamma$ in terms of power and the radiation pressure force acting on an atom. In particular, the radiation pressure force caused by Rayleigh scattering is stronger than that caused by spontaneous emission when frequency detuning is large.

### Acknowledgement

We appreciate Prof. Hakuta at The University of Electro-Communications giving us helpful comments.

### References

1. T. W. Hänsch and A. L. Schawlow: Opt. Commun. **13**, 68 (1975).
2. D. Wineland and W. M. Itano: Phys. Rev. A **20**, 1521 (1979).
3. J. Dalibard, C. Cohen-Tannoudji: J. Opt. Soc. Am. B **6**, 2023 (1989).
4. K. Shimoda: Laser, The Physical Society of Japan ed. (Maruzen 1978), p. 3 [in Japanese].
5. M. O. Scully and M. S. Zubairy: Quantum Optics (Cambridge University Press, Cambridge, 1997), Chap. 5.
6. B. R. Mollow: Phys. Rev. **188**, 1969 (1969).
7. J. L. Carlsten, A. Szöke, and M. G. Raymer: Phys. Rev. A **15**, 1029 (1977).
8. C. Cohen-Tannoudji, J. Dupont-Roc, and G. Grynberg: Atom-photon interactions (John Wiley & Sons, New York, 1992), Chaps. 2 and 6.
9. R. Loudon: The quantum theory of light (Oxford University Press, New York, 2000), 3rd ed., Chap. 8.
10. H. Walther: Adv. At. Mol. Opt. Phys. **51**, 239 (2005).
11. S. Inouye, A. P. Chikkatur, D. M. Stamper-Kurn, J. Stenger, D. E. Pritchard, and W. Ketterle: Science **285**, 571 (1999).
12. D. Schneble, Y. Torii, M. Boyd, E. W. Streed, D. E. Pritchard, and W. Ketterle: Science **300**, 475 (2003).
13. H. J. Metcalf and P. van der Straten: Laser cooling and trapping (Springer-Verlag, New York, 1999), Chaps. 2 and 3.
14. R. B. Miles, W. R. Lempert, and J. N. Forkey: Meas. Sci. Technol. **12**, R33 (2001).
15. A. Takamizawa and K. Shimoda: "Spatial intensity distribution of light under coherent interaction with an atom," submitted to arXiv.org.
16. L. Allen and J. H. Eberly: Optical resonance and two-level atoms (Dover, New York, 1987), p. 174.
17. Z. Hu and H. J. Kimble: Opt. Lett. **19**, 1888 (1994).
18. N. Schlosser, G. Reymond, I. Protsenko, and P. Grangier: Nature **411**, 1024 (2001).
19. S. Kuhr, W. Alt, D. Schrader, M. Müller, V. Gomer, and D. Meschede: Science **293**, 278 (2001).
20. A. Takamizawa, T. Steinmetz, R. Delhuille, T. W. Hänsch, and J. Reichel: Opt. Express **14**, 10976 (2006).
21. R. Grimm, M. Weidemüller, and Y. B. Ovchinnikov: Adv. At. Mol. Opt. Phys. **42**, 95 (2000).
22. V. I. Balykin, V. S. Letokhov, Y. B. Ovchinnikov, and A. I. Sidorov: Phys. Rev. Lett. **60**, 2137 (1988).
23. H. Ito, T. Nakata, K. Sakaki, M. Ohtsu, K. I. Lee, and W. Jhe: Phys. Rev. Lett. **76**, 4500 (1996).
24. M. Hammes, D. Rychtarik, B. Engeser, H. C. Nägerl, and R. Grimm: Phys. Rev. Lett. **90**, 173001 (2003).
25. A. Takamizawa, H. Ito, S. Yamada, and M. Ohtsu: Appl. Phys. Lett. **85**, 1790 (2004).